\documentclass[a4paper]{lipics-v2021}
\usepackage{hyperref}
\usepackage{xcolor,colortbl}
\usepackage{mathtools}
\usepackage{caption}
\usepackage{enumitem}
\usepackage{dsfont}
\usepackage{algorithm}
\usepackage{multicol}
\usepackage{svg}
\usepackage{bm}
\usepackage[noend]{algpseudocode}
\usepackage{amsmath}
\usepackage{breqn}
\usepackage{float}
\algnewcommand\algorithmicforeach{\textbf{for each:}}
\algnewcommand\ForEach{\item[ \algorithmicforeach]}
\algdef{S}[FOR]{ForEach}[1]{\algorithmicforeach\ #1\ \algorithmicdo}
\algtext*{EndWhile}% Remove "end while" text
\algtext*{EndIf}% Remove "end if" text
\title{Towards Statistically Significant Taxonomy Aware Co-location Pattern Detection} %TODO Please add

\titlerunning{Towards Statistically Significant Taxonomy-Aware Co-location Pattern Detection} %TODO optional, please use if title is longer than one line

\author{Subhankar Ghosh}{Department of Computer Science and Engineering, University of Minnesota}{ghosh117@umn.edu}{https://orcid.org/0000-0001-5339-1832}{}%TODO mandatory, please use full name; only 1 author per \author macro; first two parameters are mandatory, other parameters can be empty. Please provide at least the name of the affiliation and the country. The full address is optional
\author{Arun Sharma}{Department of Computer Science and Engineering, University of Minnesota}{sharm485@umn.edu}{https://orcid.org/0000-0002-6908-6960}{}
\author{Jayant Gupta}{Oracle Inc., USA}{jayant.j.gupta@oracle.com}{https://orcid.org/0000-0001-7028-9928}{}
\author{Shashi Shekhar}{Department of Computer Science and Engineering, University of Minnesota}{shekhar@umn.edu}{https://orcid.org/0000-0001-8217-3244}{}

\authorrunning{Ghosh et al.} %TODO mandatory. First: Use abbreviated first/middle names. Second (only in severe cases): Use first author plus 'et al.'

\Copyright{Authors} %TODO mandatory, please use full first names. LIPIcs license is "CC-BY";  http://creativecommons.org/licenses/by/3.0/


 %TODO mandatory; please add comma-separated list of keywords

\category{} %optional, e.g. invited paper

\relatedversion{} %optional, e.g. full version hosted on arXiv, HAL, or other respository/website
\supplement{}%optional, e.g. related research data, source code, ... hosted on a repository like zenodo, figshare, GitHub, ...

\EventEditors{John Q. Open and Joan R. Access}
\EventNoEds{2}
\EventLongTitle{42nd Conference on Very Important Topics (CVIT 2016)}
\EventShortTitle{CVIT 2016}
\EventAcronym{CVIT}
\EventYear{2016}
\EventDate{December 24--27, 2016}
\EventLocation{Little Whinging, United Kingdom}
\EventLogo{}
\SeriesVolume{42}
\ArticleNo{23}
\begin{document}
\nolinenumbers
\maketitle
\vspace*{-3.0mm}
\begin{abstract}
Given a collection of Boolean spatial feature types, their instances, a neighborhood relation (e.g., proximity), and a hierarchical taxonomy of the feature types, the goal is to find the subsets of feature types or their parents whose spatial interaction is statistically significant. This problem is for taxonomy-reliant applications such as ecology (e.g., finding new symbiotic relationships across the food chain), spatial pathology (e.g., immunotherapy for cancer), retail, etc. The problem is computationally challenging due to the exponential number of candidate co-location patterns generated by the taxonomy. Most approaches for co-location pattern detection overlook the hierarchical relationships among spatial features, and the statistical significance of the detected patterns is not always considered, leading to potential false discoveries. This paper introduces two methods for incorporating taxonomies and assessing the statistical significance of co-location patterns. The baseline approach iteratively checks the significance of co-locations between leaf nodes or their ancestors in the taxonomy. Using the Benjamini-Hochberg procedure, an advanced approach is proposed to control the false discovery rate. This approach effectively reduces the risk of false discoveries while maintaining the power to detect true co-location patterns. Experimental evaluation and case study results show the effectiveness of the approach.
\end{abstract}
\vspace*{-5.0mm}
\section{Introduction}
\label{sec:introduction}
\begin{figure}[thbp]
\centering
\includegraphics[width=0.55\textwidth]{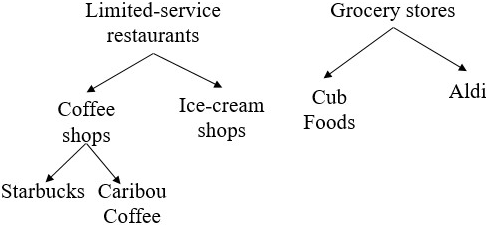}
\caption{An example taxonomy from retail.}\label{fig:TaxonomyToyExample}
\end{figure}

Given a collection of Boolean spatial feature types (which satisfy an \textit{is$-$a} property), their instances, a neighborhood relation (e.g., proximity), and a hierarchical taxonomy of the feature types, the goal is to find the subsets of feature types or their parents whose spatial interaction is statistically significant. This problem is important due to its use in taxonomy-reliant applications such as ecology (e.g., finding new symbiotic relationships across the food chain), spatial pathology (e.g., immunotherapy for cancer), retail, etc. Figure \ref{fig:TaxonomyToyExample} gives an example of a taxonomy used in retail.

Existing approaches~\cite{ghosh2022towards, ghosh2023reducing, shekhar2001discovering, huang2004discovering, li2018local, ghosh2017video} to co-location pattern detection primarily focus on identifying co-located features based on their spatial proximity and co-occurrence frequency. However, these approaches focus on a single spatial scale. In many real-world scenarios, spatial features are organized in taxonomies, where features at different levels of granularity exhibit parent-child relationships. For example, in ecological studies, species are classified into taxonomic ranks such as genus, family, and order~\cite{whittaker1972evolution}. Similarly, in urban planning, points of interest (POIs) are categorized hierarchically (e.g., food, restaurants, and fast food~\cite{yuan2012discovering}). Ignoring the hierarchical relationships among spatial features can lead to an incomplete or biased analysis of co-location patterns. If only the leaf-level features are considered, co-locations at higher taxonomy levels may be missed. Gupta and Sharma~\cite{gupta2022mining} developed the first approach for mining co-location patterns with hierarchical features. However, due to the spatial nature of the data and the multiple comparisons involved in the mining process, the method is likely to discover false positive patterns that arise by chance \cite{aldstadt2009spatial}. Traditional approaches \cite{huang2004discovering,agrawal1993mining,agrawal1994fast} on predefined thresholds such as identifying abnormal gaps \cite{sharmaphysics, sharma2022towards} or user-specified parameters to determine the significance of co-location patterns \cite{xiao2008density}. Their lack of a rigorous statistical foundation however, may lead to an inflated rate of false discoveries. Moreover, the significance of co-location patterns may vary across taxonomy levels, requiring a comprehensive analysis that considers the hierarchical structure. For example, in Figure \ref{fig:TaxonomyToyExample}, if Starbucks co-locates with Aldi in Minneapolis, that does not imply that all coffee shops (generally) co-located with Aldi in Minneapolis. We propose a novel taxonomic-aware framework for detecting statistically significant co-location patterns. Our framework incorporates the hierarchical relationships among spatial features and employs statistical techniques to control the rate of false discoveries. We introduce two approaches within this framework: (1) a baseline approach that iteratively tests the significance of co-location patterns at different levels of the taxonomy, and (2) an approach based on the \textbf{Benjamini-Hochberg} procedure \cite{benjamini1995controlling} for controlling the false discovery rate (FDR).
\vspace*{-0.5mm}

\noindent\textbf{Contributions:} 
\begin{itemize}[noitemsep]
    \vspace*{-1.0mm}
\item We formalize the problem of statistically significant taxonomy-aware co-location pattern detection and highlight its importance in various application domains.
\item We propose a Statistically Significant Taxonomy-aware co-location Miner (SSTCM) approach that  incorporates taxonomies and assesses the statistical significance of co-location patterns at different levels of the hierarchy.
\item We propose a refined FDR-based Taxonomy-aware co-location Miner (FDR-SSTCM) for controlling false discoveries using the Benjamini-Hochberg procedure.
\item We evaluate the proposed approaches using synthetic and real-world datasets, demonstrating their effectiveness of the proposed approach.
% \item We provide a comparative analysis of the two approaches, discussing their strengths, limitations, and applicability in different scenarios.
% We propose a novel framework for statistically significant taxonomy-aware co-location pattern detection to address these challenges. 
% if Starbucks and Aldi co-locate in a city (e.g., Minneapolis), that does not automatically imply that coffee shops and Aldi co-locate in Minneapolis.

% \noindent\textbf{Organization:} The paper is organized as follows. Section \ref{sec:Problem Definition} reviews the basic concepts relevant to co-location pattern detection, statistical significance, and problem formulation. Section \ref{sec:Methodology} introduces the baseline and FDR control-based approaches. Section \ref{sec:Experiments} discusses the experiment design and analysis. Section \ref{subsec:case-study} presents a case study on the Safegraph POI dataset. Finally, Section \ref{sec:related} discusses the related works while Section \ref{sec:Conclusion} concludes the paper and outlines future research directions.
\end{itemize}
\vspace*{-1.0mm}
\section{Basic Concepts and Problem Definition}
\label{sec:Problem Definition}
% First, we review basic concepts related to co-location detection and statistical significance testing. Then, we formally define statistically significant taxonomy-aware co-location pattern detection.
\vspace*{-1.0mm}
% \subsection{co-location detection:}
% \label{sec:colocation_detection}
A \textbf{feature instance} is a geo-located spatial entity which is a type of Boolean feature $f$ with a geo-reference point location $p$ (e.g., latitude, longitude), represented as $<f,p>$. Multiple instances of a feature are represented as $f_{i}$ and can be related to other feature instances $f_{j}$ via a \textbf{neighbor relation} $\mathcal{R}$. For example, geographic proximity is represented as $\mathcal{R}_{f_{i}, f_{j}}$ $\leq$ $\theta$, where $\theta$ is the neighbor relation threshold. In a \textbf{neighbor graph}, we represent features that satisfy such relations as a \textit{node} and their relationship as an $edge$.

A \textbf{co-location candidate} $C$ is a set of features defined in a given study area ($SA$) or a sub-region ($r_{g}$) where $r_{g}$ $\in$ $SA$. An instance of a \textbf{co-location} satisfies the neighborhood relation $\mathcal{R}$ and forms a \textbf{clique}. \textbf{Co-location patterns} \cite{shekhar2001discovering} are the set of prevalent co-location candidates (based on a prevalence measure, e.g., $pi$), i.e., candidates comprised of features having a high positive spatial interaction.
\noindent
A \textbf{participation ratio ($pr$)} is the ratio of feature instances participating in a relation $\mathcal{R}$ to the total number of instances inside the study region $(SA)$. A \textbf{participation index ($pi$)} is the minimal participation ratio of all feature types in a co-location candidate. 
A \textbf{taxonomy ($T$)} of feature-types represents an ‘is\_a’ relation between two boolean feature types (e.g., Starbucks is\_a Coffee Shop).
% The participation index quantifies the spatial interaction within features.
% For a given co-location candidate $C$ and feature $f$, it is represented as $pr(f, C)$ as shown in Equation \ref{equation:PR}:
% \begin{equation}
% % \centering
%     \label{equation:PR}
%     pr(f, C) = \frac{participating\_instances(f, C)}{instance(f)}
% \end{equation}

% as described in Equation \ref{equation:PI_basic_concepts}:
% \begin{equation}
%     \label{equation:PI_basic_concepts}
%     % https://tex.stackexchange.com/a/130517
%     pi(C) = \underset{f \in C} {min}(pr(f, C))
% \end{equation}
% \subsection{Regional-co-location Pattern}
% \emph{pr(C, f_{i})} = $\frac{number of objects of feature in relation}{total number objects of feature}$
% \vspace*{-5.0mm}
% \subsection{Statistical Significance in co-location Detection:}
% \label{sec:stat_sig_test_basic_concepts} 
% A statistically significant co-location determines whether an assigned positive spatial interaction between features is statistically significant or could have been observed if the features were in complete spatial randomness (CSR). 
A \textbf{statistically significant co-location} determines whether an observed positive spatial interaction between features is genuine or could have occurred under complete spatial randomness (CSR). A \textbf{statistical significance test} for a co-location pattern assesses the probability of observed results if the features were spatially independent (null hypothesis).

The \textbf{multiple comparisons problem} \cite{rupert2012simultaneous} arises when multiple inferences are made simultaneously, increasing the likelihood of incorrectly rejecting the null hypothesis. Addressing this issue often requires stricter significance thresholds for individual comparisons. The \textbf{Benjamini-Hochberg procedure} \cite{benjamini1995controlling} is used for controlling our false discovery rate (FDR), which is the expected proportion of false positives among all significant hypotheses. \\(See the appendix for more definitions.)
\vspace*{-1.0mm}
\textbf{Formal Problem Definition}: Given a set \(F\) of spatial features, a significance level \(\alpha\), FDR control level \(q\), and a neighbor relationship \(\mathcal{R}\) in a taxonomy tree \(T\), we aim to find statistically significant taxonomy-aware co-location patterns, \(C\). Our objective is to reduce false discoveries, focusing on patterns of smaller size and higher statistical confidence.

% \vspace*{-5.0mm}
% \subsection{Formal problem formulation}
% \label{sec:problem_definition}
% \vspace*{-3.0mm}
% The problem of statistically significant taxonomy-aware co-location pattern detection is as follows:\newline
% \textbf{Input:}
% \begin{enumerate}[noitemsep,topsep=2pt]
%     \vspace*{-0.5mm}
%     \item A set ($F$) of spatial features in a study area
%     % \item $N$ geo-located spatial feature instances.
%     \item A statistical significance level $\alpha$.
%     \item An FDR control level $q$.
%     \item A neighbor relationship ($\mathcal{R}$).
%     \item A Taxonomy tree $T$
%     \vspace*{-0.5mm}
% \end{enumerate}
% \textbf{Output:} Statistically significant taxonomy-aware co-location patterns, $C$ where $C\subset F$.
% \\\textbf{Objective:} Reducing false discoveries.\\
% \textbf{Constraints:} Higher statistical confidence of output patterns. Patterns of smaller size due to high computation cost.
% Traditional regional-co-location miners are data-driven.

\textbf{Reasoning behind problem output:} Testing for statistical significance in taxonomy-aware co-location analysis reduces the identification of spurious patterns, which may arise from class imbalance or spatial auto-correlation. High-cardinality features (e.g., \( f_B \) in \ref{subfig:TaxonomyToyExample2}) pose multiple comparisons problems, increasing the false discovery rate. Setting a specific \( \alpha \) level is not enough; controlling the overall false discovery rate is necessary.

\section{Methodology}
\label{sec:Methodology}
\vspace*{-1.0mm}
\subsection{Statistically Significant Taxonomy-aware Co-location Miner}
\label{sec:baseline}
\vspace*{-1.0mm}

In our study, we utilize a tree-like structure (taxonomy) to represent the hierarchical relationships among spatial features, where each node corresponds to a feature type and edges denote parent-child relationships. The most specific feature types are represented by leaf nodes, while internal nodes signify broader categories. An example of this taxonomy in a retail dataset is shown in Figure \ref{fig:TaxonomyToyExample}. To detect co-location patterns, our baseline approach initially identifies patterns at the leaf level, calculating co-location strength using metrics such as the participation index \cite{huang2004discovering}. A pattern is considered significant if its strength exceeds a predetermined threshold, which may be set using domain knowledge or statistical methods like randomization tests \cite{besag1977simple}. For patterns involving leaf-level features, such as $f_{C}$ and $f_{H}$, we use Algorithm \ref{Appendix:algorithm1} (in Appendix) to compare observed and null hypothesis datasets to compute a $p$-value, determining statistical significance against a threshold $\alpha$. Additionally, for patterns involving non-leaf features with children, such as $f_{B}$ and $f_{H}$, a post-order traversal is performed to check the significance of interactions between the children nodes ($f_{D}$, $f_{E}$, and $f_{F}$) and the leaf node $f_{H}$. If they are significant, this implies the co-location between the parent node $f_{B}$ and the leaf node $f_{H}$ is also significant. Our approach systematically explores co-location patterns at varying levels of granularity, integrating the taxonomy's hierarchical structure into the analysis process.

\textbf{Limitations and Challenges}: While our baseline method effectively incorporates taxonomies into co-location pattern detection, it does present certain limitations and challenges. One significant issue is the multiple comparisons problem, where the iterative testing across various taxonomy levels increases the risk of Type I errors (false discoveries) due to the heightened number of tests, which boost the likelihood of significant results occurring by chance. Furthermore, the baseline algorithm requires each child node feature of an intermediate taxonomy tree to demonstrate a statistically significant co-location with other features in the candidate pattern, a stringent criterion that tends to elevate the rate of Type II errors (false negatives). To overcome these challenges, we propose an advanced approach that includes specific techniques aimed at controlling the false discovery rates inherent in the taxonomy-aware co-location pattern detection process. More details shown in Algorithm \ref{algorithm2} (in Appendix).

% While the baseline approach provides a straightforward way to incorporate taxonomies into co-location pattern detection, it has some limitations and challenges:

% \textbf{Multiple Comparisons Problem}: The iterative significance testing at different levels of the taxonomies can lead to an increased risk of false discoveries (Type I errors) due to multiple comparisons. As the number of tests increases, the chance of obtaining significant results by random chance also increases.

% \textbf{Higher False negatives}: The baseline algorithm necessitates every children's feature node of an intermediate taxonomy tree to have a statistically significant co-location with the other features in the candidate pattern. This strict method can lead to higher false negatives or Type-$II$ errors.

% To address these limitations and challenges, we propose an advanced approach incorporating techniques for controlling false discoveries of the taxonomy-aware co-location pattern detection process.
\vspace*{-1.0mm}
\vspace*{-1.0mm}
\subsection{FDR-based Taxonomy-aware co-location Miner}
\label{sec:proposed2}
While the baseline $SSTCM$ algorithm can reduce spurious pattern detection, it may be overly conservative in some situations, reducing its statistical power. To strike a balance between controlling false discoveries and maintaining the ability to detect true co-location patterns, we propose a second approach that incorporates the Benjamini-Hochberg procedure \cite{benjamini1995controlling} for controlling the false discovery rate (FDR). The Appendix \ref{appendix} provides more details on the procedure.

\textbf{False Discovery Rate and Benjamini-Hochberg Procedure}: The false discovery rate (FDR) is the expected proportion of false positives among all significant hypotheses. Controlling the FDR is a less stringent criterion than the baseline, as it allows for a certain proportion of false positives while focusing on the overall rate of false discoveries.

The steps for the Benjamini-Hochberg procedure are as follows:
\begin{enumerate}[noitemsep,topsep=2pt]
\vspace*{-0.5mm}
\item Order the p-values of the m hypothesis tests from smallest to largest: p(1), p(2), ..., p(m).
\item Set the desired FDR level q (e.g., 0.05).
\item Find the largest integer k such that $p(k) \leq (k/m) * q$.
\item Reject the null hypotheses for all tests with p-values smaller than or equal to p(k).

By controlling the FDR, the Benjamini-Hochberg procedure ensures that, on average, no more than a specified proportion q of the rejected hypotheses are false positives.
\end{enumerate}
\begin{algorithm}[thp]
\caption{FDR-based Taxonomy-aware co-location Miner ($FDR$\textendash$SSTCM$) snippet} \label{algorithm3}
\footnotesize
\begin{flushleft}
\vspace{-10pt}
\hspace{\algorithmicindent}\textbf{Input:}\\
\hspace*{\algorithmicindent} - A spatial dataset \emph{S} consisting of features \{$f_{A}, f_{B}, ...$\}\\
\hspace*{\algorithmicindent} - Distance threshold $d$ (in meters)\\
\hspace*{\algorithmicindent} - Taxonomy tree $T$\\
\hspace*{\algorithmicindent} - FDR control level $q$\\
\hspace*{\algorithmicindent} - Statistical significance level $\alpha$\\
\hspace*{\algorithmicindent} - Candidate pattern set ${C}$\\
\hspace*{\algorithmicindent}\textbf{Output:}\\
\hspace*{\algorithmicindent} A statistically significant subset of co-location patterns $C_s \subset {C}$ and their p-values ($p$-$value_{C_s}$)\\
% which is a subset of features \{$f_{A}, f_{B}, ...$\} from \emph{S}
% \hspace*{\algorithmicindent}\textbf{Variables:}\\
% \hspace*{\algorithmicindent} Distance between feature instances $d$ 
% d is always bounded, where the bounds are determined empiricially from the given observed data
\end{flushleft}
\vspace{-10pt}
\begin{algorithmic}[1]
    \Procedure{FDR\textendash SSTCM}{}
        \State{$\vdots$}
        \setcounter{ALG@line}{21}
    \EndProcedure

    \Procedure{Traversal}{$f_p, f_l$}
        \State{$\vdots$}
        \setcounter{ALG@line}{28}
    \EndProcedure

    \Procedure{PostOrder}{$f_p, f_l$}
        % \State {$result$ $\leftarrow$ $True$}
         \ForEach{$f_{c} \in children(f_{p})$}
            \State {$result, p-value = Traversal(f_c, f_l)$}
            \State {$p-value_{list}$ $\leftarrow$ append $p-value$}
         \EndFor
         \State {Sort $p-value_{list}$ in ascending order $p_{(1)}, p_{(2)}, ..., p_{(m)}$}
         \State {$L = max\{j: p_{(j)} < qj/m\}$} \Comment{Benjamini-Hochberg procedure}
         \If{$L \leq 2$}
            \State{return $False$}
        \EndIf
         \State {$f_p^{'}$ $\leftarrow$ Update $f_p$ with it's children instances}
         \State{return $True$}
    \EndProcedure
\end{algorithmic}
\end{algorithm}
\setlength{\textfloatsep}{5pt}

\begin{figure*}[h]
\centering
\includegraphics[width = 4.2cm]{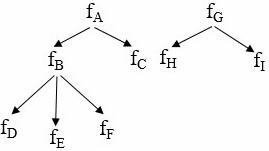}\hfill
\caption{An example taxonomy.}
\label{subfig:TaxonomyToyExample2}
\end{figure*}

\textbf{Incorporating the Benjamini-Hochberg Procedure in Taxonomy-Aware Co-location Pattern Detection}: In $FDR$–$SSTCM$, the Benjamini-Hochberg procedure is used at each level of the taxonomy when some of the constituent features of the candidate pattern are intermediate nodes in the taxonomy tree. As shown in Figure \ref{subfig:TaxonomyToyExample2}, when checking for the significance of the pattern ($f_B, f_H$), the $FDR$–$SSTCM$ algorithm checks for the significance of the children of $f_B$ with $f_H$. Applying the Benjamini-Hochberg procedure in this step ensures that at least a few children of $f_B$ have a significant co-location with $f_H$ before we conclude that ($f_B, f_H$) is a statistically significant co-location pattern.

% \textbf{Advantages and Limitations}:

The proposed approach using the Benjamini-Hochberg procedure has several advantages:
\begin{enumerate}[noitemsep,topsep=2pt]
\vspace*{-0.5mm}
\item \textbf{FDR Control:} The approach controls the FDR and provides a means to balance between minimizing false discoveries and maintaining statistical power, allowing a manageable proportion of false positives to focus on the overall false discovery rate.
\item \textbf{Increased Power:} Compared to the baseline or family-wise error rate (FWER) control methods like the Holm-Bonferroni method, the Benjamini-Hochberg procedure generally has higher statistical power, meaning it is more likely to detect true co-location patterns.
\item \textbf{Adaptability:} The approach can be applied to a variety of co-location strength measures and significance tests, making it suitable for different datasets and problem settings.
\end{enumerate}
However, there are also some limitations to consider:
\begin{enumerate}[noitemsep,topsep=2pt]
\vspace*{-0.5mm}
\item \textbf{Assumption of Independence:} The Benjamini-Hochberg procedure can inflate FDR levels if its assumption of test independence or positive dependence is violated.
\item \textbf{Choice of FDR Level:} Specifying the FDR level q is crucial as it influences results, requiring domain knowledge or alignment with application needs. While the Benjamini-Hochberg procedure effectively manages FDR, balancing false and true positives, the results generated demand careful interpretation and possible further validation. Despite the potential for false positives, it remains a valuable alternative to baseline methods in detecting significant taxonomy-aware co-location patterns.
% While the FDR control balances false positives and true positives, it is important to interpret the results carefully. The rejected hypotheses may still contain some false positives, and further validation or domain expertise may be necessary to confirm the significance of the detected co-location patterns.
% Despite these limitations, the proposed approach using the Benjamini-Hochberg procedure offers a powerful and flexible method for controlling the FDR in statistically significant taxonomy-aware co-location pattern detection. It provides a valuable alternative to the baseline method, particularly when a certain level of false positives is acceptable and the focus is on the overall rate of false discoveries.
\end{enumerate}

\section{Experimental Evaluation}
\label{sec:Experiments}
\begin{figure*}[h]
\centering
{\includegraphics[width = 9.5cm]{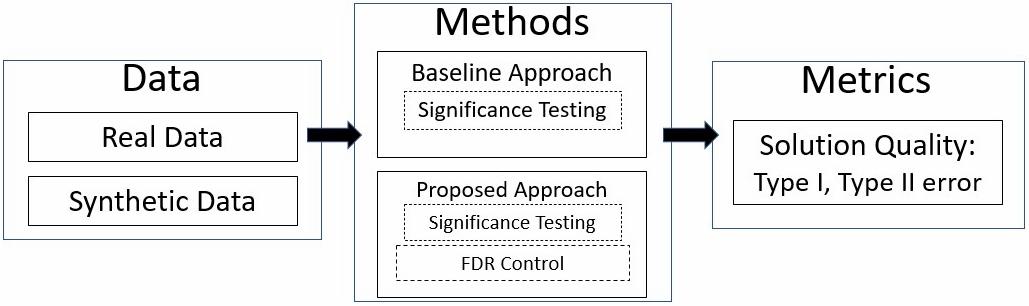}}\hfill
\caption{Experiment design.}
\label{subfig:validation_framework}
\end{figure*}

Our experimental goal was to compare solution quality between $SSTCM$ and $FDR$-$SSTCM$ and Figure \ref{subfig:validation_framework} shows the experiment design.

We performed controlled experiments on synthetic data to compare the solution quality of $SSTCM$ with $FDR$-$SSTCM$. The metrics for comparison were $Type$-$I$ and $Type$-$II$ error rates.
$P($Type-I$\ error) = P(Reject\ H_{0} | H_{0}\ is\ true)$ represents the probability of incorrectly rejecting the null hypothesis $H_{0}$, while $P($Type-II$\ error) = P(Fail\ to\ reject\ H_{0} | H_{0}\ is\ false)$ indicates the likelihood of failing to reject a false $H_{0}$. Table \ref{tab:sol_qual} displays the experiment results: FDR-SSTCM has a similar Type-I error rate but a significantly lower Type-II error rate compared to SSTCM. The baseline SSTCM algorithm mandates each child intermediate nodes within a taxonomy tree to co-locate with other features to consider the parent node's significance resulting in increased Type-II errors. In addition, unlike FDR-SSTCM, SSTCM does not address the multiple comparisons issue and may exacerbate Type-I errors in the presence of class imbalances.
% \noindent $P($Type-I$\ error)=P(Reject\ H_{0} | H_{0}\ is\ true)$, where $H_{0}$ represents the Null Hypothesis. $P($Type-II$\ error)$=$P(Fail\ to\ reject\ H_{0} | H_{0}\ is\ false)$. Table \ref{tab:sol_qual} shows the experiment results. FDR-SSTCM exhibits a comparable $Type$-$I$ error rate and significantly lower $Type$-$II$ error rate than $SSTCM$. Baseline $SSTCM$ requires every child of an intermediate node (in a taxonomy tree) to co-locate with the other features in a candidate pattern before it checks for the significance of the parent node with the other features. This strict procedure can produce higher false negatives ($Type$-$II$ error rate). The baseline also does not handle the multiple comparisons problem. Also, in case of a class imbalance problem, $SSTCM$ can lead to higher Type-I error than $FDR$-$SSTCM$.

\begin{table}[tbhp]
\centering
{
\footnotesize
\caption{$FDR$-$SSTCM$ has lower $Type$-$I$ error rate and much lower $Type$-$II$ error rate as compared baseline $SSTCM$}\label{tab:sol_qual} 
\begin{tabular}{|p{0.95cm}|p{2.3cm}|p{2.3cm}||p{2.4cm}|p{2.4cm}|}
 \hline
%   & \multicolumn{2}{c|}{\textbf{False positive rate}}\\ \hline
 \textbf{Pattern} & \textbf{$SSTCM$ $Type$-$I$} & \textbf{$FDR$-$SSTCM$ $Type$-$I$} & \textbf{$SSTCM$ $Type$-$II$} & \textbf{$FDR$-$SSTCM$ $Type$-$II$} \\ \hline
 {$f_A$, $f_G$} & 0.05 & 0.03 & 0.35 & 0.21\\ \hline
 {$f_E$, $f_H$} & 0.02 & 0.02 & 0.2 & 0.13\\\hline
 {$f_B$, $f_H$} & 0.04 & 0.03 & 0.27& 0.16\\ \hline
\end{tabular}
}
\end{table}

% \section{Case Study}
% \label{subsec:case-study}
\textbf{Case Study:} We extended our previous case study \cite{ghosh2022towards} to demonstrate the effectiveness of the proposed approach. The dataset, provided by SafeGraph---a vendor of mobility data---offers anonymized and aggregated location data to researchers studying the impact of COVID-19 on movement patterns towards various Points of Interest (POIs) across 1,473 retail brands in Minnesota. The data organization follows the hierarchical North American Industry Classification System (NAICS), which classifies businesses from level 1 to level 5 (e.g., if NAICS level 1 is 'Retail Trade', then more detailed information is fleshed out at subsequent levels, from level 2 to level 5). In our case study, we found that the co-location of limited-service restaurants, specifically Subway and McDonald's, and coffee shops like Starbucks, was statistically significant within approximately 1400 meters in Hennepin County. Further global searches confirmed similar significant co-location patterns for Fast Food Restaurants and Coffee Shops, and other category pairings like Starbucks with Olive Garden showed significance, though this pattern did not hold for general categories of coffee shops and full-service restaurants.

\section{Related Work}
\label{sec:related}
Co-location pattern detection is used to discover subsets of spatial features that frequently co-occur in close proximity~\cite{shekhar2001discovering} and has been extensively studied in the spatial data mining literature. Huang et al.~\cite{huang2004discovering} introduced the concept of spatial co-location patterns and proposed a framework for mining such patterns using a spatial join-based approach. Yoo et al.~\cite{yoo2006joinless} developed a joinless approach for mining co-location patterns, which improves computational efficiency by avoiding the need for spatial joins. These early works focused on discovering co-location patterns at a single spatial scale without considering the hierarchical relationships among spatial features. Gupta and Sharma \cite{gupta2022mining} explored the detection of co-location patterns within hierarchical relationships. However, their approach did not consider statistical significance, leading to spurious patterns, especially in the presence of class imbalance among child nodes in a taxonomy.
% Xiao et al. \cite{xiao2008density} proposed a density-based approach for discovering co-location patterns in spatial data with multiple resolutions. They defined the concept of density-based co-locations and developed an algorithm to detect such patterns across different spatial scales. However, their approach did not explicitly consider the taxonomic relationships among spatial features.
% Huang et al. \cite{huang2006relationships} introduced the concept of regional co-location patterns and proposed a method for discovering such patterns using a spatial clustering approach. They defined regions based on the density of spatial features and discovered co-location patterns within these regions. While this approach considered the spatial heterogeneity of co-location patterns, it did not consider the hierarchical relationships among features.
% Several studies have recognized the importance of statistical significance in co-location pattern detection. Huang et al. \cite{huang2006mining} proposed the cross-K function to assess the significance of co-location patterns. They used Monte Carlo simulations to generate random spatial point patterns and compared the observed co-location patterns against these simulated patterns to determine their statistical significance. However, their approach did not consider the multiple testing problem when evaluating the significance of numerous co-location patterns.
Barua et al. \cite{barua2013mining} introduced the concept of statistical significance in global co-location pattern detection, and Ghosh et al.~\cite{ghosh2022towards} introduced statistical significance in regional co-location pattern detection, but their work did not consider taxonomies or the multiple comparisons problem. Ghosh et al. \cite{ghosh2023reducing} addressed the multiple testing problem in the context of co-location pattern detection. Using the Bonferroni Correction, they proposed a method for controlling the family-wise error rate (FWER). However, they did not consider the hierarchical relationships among spatial features. Recently, deep learning-based spatial association mining~\cite{farhadloo2022samcnet, farhadloo2023spatial, farhadloo2024towards} has been proposed, but none of them considered the hierarchical relationship between features. Spatial association has also been studied in various other domains~\cite{ghosh2023reducingiguide, ghosh2017video, li2022cscd, yang2023data} where considering the hierarchical relationship might be valuable.
\vspace*{-1.0mm}
\section{Conclusion and Future Work}
\label{sec:Conclusion}
In this paper, we propose a statistically significant taxonomy-aware co-location miner (SSTCM) and refine the problem of statistical significance with the proposed FDR-based taxonomy-aware co-location miner, utilizing the Benjamini-Hochberg procedure. We evaluate our approaches using synthetic and real-world datasets, demonstrating their effectiveness. Finally, we provide a comparative analysis of the two approaches, discussing their strengths, limitations, and provide a case study on retail establishments in Minnesota using the proposed approach.\\
\noindent \textbf{Future Work:} We plan to explore other methods to reduce Type-I errors (false positives) further while also addressing Type-II errors (false negatives) and further providing computational efficiency to handle patterns of a large number of features. Finally, we plan to add a temporal dimension to these patterns.

\bibliography{document}
% \pagebreak
\newpage
\appendix
\vspace*{-1.0mm}
\section{Appendix}
\label{appendix}
A \textbf{null hypothesis} $(H_{0})$ posits no spatial interaction between dataset features, suggesting their independence. Conversely, an \textbf{alternative hypothesis} $(H_{a})$ asserts positive spatial interaction among features in a specified region, challenging $H_{0}$.

\noindent A \textbf{Type-I error} refers to the erroneous rejection of an actually true null hypothesis (or a false positive) while a \textbf{Type-II error} refers to the failure to reject a null hypothesis $(H_{0})$ that is actually false (or a false negative).

\noindent\textbf{Null hypothesis generation}: 

\begin{itemize}[noitemsep,topsep=1pt]
    \item For an identical distribution, we generate the same number of feature instances across the study area using summary statistics, ensuring that the null hypotheses datasets closely model the observed dataset in each partition.
    \item We sample instances using a Poisson point process to ensure independence, analyzing them with a pair correlation function (PCF) up to a data-driven distance $d$; $g(d)>1$ indicates clustering, while $g(d)=1$ signifies complete spatial randomness (CSR).
\end{itemize}

\noindent \textbf{Statistical significance test}: The participation index ($pi$) quantifies spatial interaction strength; we compute the probability that the observed data's $pi$ for a pattern $C$, represented as $pi_{obs}(C)$, differs from its null hypothesis index, $pi_{\emptyset}(C)$.
\begin{equation}
\label{PIObs}
    p = pr(pi_{\emptyset}(C) \geq pi_{obs}(C)) = \frac{R^{\geq pi_{obs}} + 1}{R + 1},
\end{equation}
where $R^{\geq pi_{obs}}$ represents the number of Monte Carlo simulations within which the participation index ($pi_{\emptyset}(C)$) for pattern $C$ is greater than in the observed data ($pi_{obs}(C)$) and $R$ refers to the total number of Monte Carlo simulations. If $p \leq \alpha$, we consider $pi_{obs}(C)$ as statistically significant at level $\alpha$.

The \textbf{Benjamini-Hochberg} procedure \cite{benjamini1995controlling} is a widely used method for controlling the false discovery rate (FDR) in multiple-hypothesis testing. The FDR is the expected proportion of false positives among all significant hypotheses. Unlike family-wise error rate (FWER) control methods, such as the Bonferroni correction, which controls the probability of making at least one false discovery, the Benjamini-Hochberg procedure focuses on the proportion of false discoveries among all rejected hypotheses. It provides a more powerful and less conservative approach to multiple testing by allowing a certain level of false positives while ensuring that the FDR is controlled at a desired level. The procedure works by ranking the p-values from smallest to largest and comparing each p-value to a threshold that depends on its rank and the desired FDR level. The Benjamini-Hochberg procedure has been widely applied in various fields, including genomics, neuroscience, and spatial data mining, where multiple comparisons are common, and controlling the FDR is crucial for drawing meaningful conclusions from the data.

\begin{algorithm}[thp]
\caption{Significance testing} \label{Appendix:algorithm1}
\footnotesize
\begin{flushleft}
\vspace{-10pt}
\hspace{\algorithmicindent}\textbf{Input:}\\
\hspace*{\algorithmicindent} - A spatial dataset \emph{S} consisting of features \{$f_{A}, f_{B}, ...$\} in a study area \\
\hspace*{\algorithmicindent} - Statistical significance level $\alpha$ and candidate co-location pattern $C$\\
% \hspace*{\algorithmicindent} - A candidate co-location pattern $C$\\
\hspace*{\algorithmicindent} - A set of $R$ Null hypotheses ($NH_{\emptyset}$) data each modelled as co-location $C$\\
% \hspace*{\algorithmicindent} - Distance $d$ for participation index $(pi)$ calculation\\
\hspace*{\algorithmicindent}\textbf{Output:}\\
\hspace*{\algorithmicindent} 1. $C$ is significant or not \\
% which is a subset of features \{$f_{A}, f_{B}, ...$\} from \emph{S}
% \hspace*{\algorithmicindent} $pi_{obs}$ of $C$ in $r_{g}$\\
 \hspace*{\algorithmicindent} 2. $p$-$value_C$\\
\end{flushleft}
\vspace{-10pt}
\begin{algorithmic}[1]
    \Procedure{Significance Testing}{}
    \State{Statistically significant result $SSR_C$ $\leftarrow$ False}
    % \State{R := Number of null hypotheses}
            \State {Counter $R^{\geq pi_{obs}}$ $\leftarrow$ 0}
            \State {Calculate $pi_{obs}$ for $C$ at $d$}
            \For{$i \in$ [$1,R$]}
                \State{Calculate the $pi_{\emptyset, i}$ of $C$ at $d$ in the $i^{th}$ $NH_{\emptyset}$}
                \If{$pi_{\emptyset, i} \geq pi_{obs}$}
                    \State $R^{\geq pi_{obs}} \leftarrow R^{\geq pi_{obs}} + 1$
                \EndIf
            \EndFor
                \State $p$-$value_{C}= \frac{R^{\geq pi_{obs}}+1}{R+1}$
                \If{$p$-$value_{C}$ $\leq$ $\alpha$}
                    \State {$SSR_C$ $\leftarrow$ True \Comment{(i.e., $C$ is statistically significant)}}
                    % \State $significance\_result$ $:=$ Significant
                \Else
                    \State{$SSR_C$ $\leftarrow$ False \Comment{(i.e., $C$ is not statistically significant)}}
                    % \State $significance\_result$ $:=$ Not Significant
                \EndIf
    \State {return $SSR_C$, $p$-$value_{C}$}
\EndProcedure
\end{algorithmic}
\end{algorithm}
\setlength{\textfloatsep}{5pt}

\begin{algorithm}[thp]
\caption{Statistically Significant Taxonomy-aware co-location Miner ($SSTCM$)} \label{algorithm2}
\footnotesize
\begin{flushleft}
\vspace{-10pt}
\hspace{\algorithmicindent}\textbf{Input:}\\
\hspace*{\algorithmicindent} - A Spatial dataset \emph{S} consisting of features \{$f_{A}, f_{B}, ...$\}\\
% \hspace*{\algorithmicindent} - Distance threshold $d$ (in meters)\\
\hspace*{\algorithmicindent} - Taxonomy tree $T$\\
\hspace*{\algorithmicindent} - Statistical significance level $\alpha$\\
\hspace*{\algorithmicindent} - Candidate pattern set ${C}$\\
\hspace*{\algorithmicindent}\textbf{Output:}\\
\hspace*{\algorithmicindent} Subset of co-location patterns $C_s \subset {C}$ which are statistically significant and their p-values ($p$-$value_{C_s}$)\\
% which is a subset of features \{$f_{A}, f_{B}, ...$\} from \emph{S}
% \hspace*{\algorithmicindent}\textbf{Variables:}\\
% \hspace*{\algorithmicindent} Distance between feature instances $d$ 
% d is always bounded, where the bounds are determined empirically from the given observed data
\end{flushleft}
\vspace{-10pt}
\begin{algorithmic}[1]
    \Procedure{Statistically Significant Taxonomy-aware co-location Miner}{}
    \ForEach{$f_{k}$ in \{$f_{A}, f_{B}, ...$\}}
    \State {Generate $R$ null hypotheses ($NH_{\emptyset}$) using summary statistics}
    \EndFor
    \ForEach{candidate pattern $C_{m} \in \{C_{1}, C_{2},...,C_{M}\}$}
        \State{$f_1, f_2 \in C_m$}
        \If{$f_1, f_2$ are leaf nodes in $T$:}
            % \State{$SSR_{C_m}$, $p-value$ $\leftarrow$ Significance Testing(S, $\alpha$, $C_{m}$, $NH_{\emptyset}$)}
            \State{return Significance Testing(S, $\alpha$, $C_{m}$, $NH_{\emptyset}$)}
        \ElsIf{$f_1$ is leaf node and $f_2$ is a parent node in $T$:}
            % \State{$SSR_{C_m}$, $p-value$ $\leftarrow$ Traversal($f_2, f_1$)}
            \State{return Traversal($f_2, f_1$)}
        % \ElsIf{$f_2$ is leaf node and $f_1$ is a parent node in $T$:}
        %     \State{$SSR_{C_m}$, $p-value$ $\leftarrow$ Traversal($f_1, f_2$)}
        %     \State{return $SSR_{C_m}$, $p-value$}
        \Else
            \State{Replace $f_2$ with it's children instances}
            \State{$SSR_{C_m}^{1}$, $p-value_1$ $\leftarrow$ Traversal($f_1, f_2$)}
            \State{Replace $f_1$ with it's children instances}
            \State{$SSR_{C_m}^{2}$, $p-value_2$ $\leftarrow$ Traversal($f_2, f_1$)}
            \If{$SSR_{C_m}^{1}$ and $SSR_{C_m}^{2}$ are both $True$:}
                \State{return Significance Testing(S, $\alpha$, $C_{m}$, $NH_{\emptyset}$)}
            \EndIf
        \EndIf
    \EndFor
    \EndProcedure

    \Procedure{Traversal}{$f_p, f_l$}
        \If{$f_p$ is a leaf node in $T$:}
            \State{return Significance Testing(S, $\alpha$, $(f_p, f_l)$, $NH_{\emptyset}$)}
        \EndIf
        $post$-$order$-$result$ = PostOrder(f_p, f_l)
        \If{$post$-$order$-$result$ is $True$}
            \State{return Significance Testing(S, $\alpha$, $(f_p^{'}, f_l)$, $NH_{\emptyset}$)}
        \Else
            \State{return $False, 1$}
        \EndIf
    \EndProcedure

    \Procedure{PostOrder}{$f_p, f_l$}
        \State {$result$ $\leftarrow$ $True$}
         \ForEach{$f_{c} \in children(f_{p})$}
            \State {$result, p-value = Traversal(f_c, f_l)$}
            \If{$result$ is $False$}
                \State {return $result, 1$}
            \EndIf
         \EndFor
         \State {$f_p^{'}$ $\leftarrow$ Update $f_p$ with it's children instances}
    \EndProcedure
\end{algorithmic}
\end{algorithm}
\setlength{\textfloatsep}{5pt}

% \textbf{Synthetic data generation:} We began with a space partitioning $(R_{g})$, a maximum union (or traversal) of regions $(L_{max})$, and a number of regional-co-location patterns i.e., pairs of $<r_{g},C>$. We then generated reference points within the partitions using the Poisson point process. At each reference point, we generated circles of diameter $d_g$ which was determined empirically for each region in $R_{g}$ in the observed dataset. The diameter signifies the smallest distance between features in a co-location $C$ at which they become statistically significant regional co-locations. We populated each
% circle with instances of $C$. We note that the circles were only used to place colocated instances in a region and were not separate partitioning. Figure \ref{fig:syn_data_gen} shows the synthetic data generation process.

% \begin{figure}[htbp]
%   \centering
%   \graphicspath{ {./images/} }
%   \includegraphics[width=0.60\linewidth]{images/syn_data.png}
%   \caption{Synthetic data generation process.}
%   \label{fig:syn_data_gen}
% \end{figure}

% \label{Appendix:Experiment_design}
% \textbf{Experiment design:} Figure \ref{subfig:validation_framework} shows the overall validation framework. The metric for comparing the solution quality of $SSRCM$ with MultComp-RCM was the false positive rate (FPR), while the runtime comparisons were based on the execution time (in seconds) of the individual algorithms. The experiments were done on real (Safegraph POI) and synthetic data to perform comparative and sensitivity analyses.

\end{document}